DISCOVERING ATOMISTIC PATHWAYS FOR SUPPLY OF METAL ATOMS TO GRAPHENE SURFACE FROM METHYL-BASED PRECURSORS

*Davide G. Sangiovanni,*[1] *Ricardo Faccio,* [2] *Gueorgui K. Gueorguiev,*[1] *and Anelia Kakanakova-Georgieva*[1] *

[a] Department of Physics, Chemistry and Biology (IFM), Linköping University, 581 83 Linköping, Sweden

[b] Área Física & Centro Nanomat, DETEMA, Facultad de Química, Universidad de la República, Av. Gral. Flores 2124, C.P. 11800, Montevideo, Uruguay

* E-mail: anelia.kakanakova@liu.se

Conceptual 2D group III nitrides and oxides (e.g., 2D InN and 2D InO) in heterostructures with graphene have been realized by metalorganic chemical vapor deposition (MOCVD). MOCVD is credited with being central to fabrication of established semiconductor materials and by purpose for an advance in emergent semiconductor materials at the 2D limit. A defining characteristic of MOCVD is the employment of metalorganic precursors such as trimethyl-indium, -gallium, and -aluminum, which contain (strong) metal-carbon bonds. Mechanisms that regulate MOCVD processes at the atomic scale are largely unknown. Here, we employ density-functional molecular dynamics – accounting for van der Waals interactions – to locate reaction pathways responsible for dissociation of trimethylindium (TMIn) precursor in the gas phase as well as on top-layer and zero-layer graphene. The simulations reveal how collisions with hydrogen molecules, intramolecular or surface-mediated proton transfer, and direct TMIn/graphene reactions assist TMIn transformations, which ultimately enables delivery of In monomers, or InH and $CH_3In$ admolecules, on graphene. Results presented also show how TMIn + $H_2$ reactions on graphene, or in the gas phase, lead to formation of methane, ethane, propane, ethene hydrocarbons, and atomic hydrogen. This work provides knowledge for understanding thin-film nucleation and intercalation mechanisms at the atomic scale and for overcoming challenges in integration of 2D materials and graphene heterostructures in technology.

## 1. INTRODUCTION

Exploration and development of materials exhibiting new compositions and/or unusual structural and electronic properties in ever-scaling-down dimensionality, is a "hot" research topic undergoing continuous update. Group III nitrides (AlN, GaN, and InN), for example, have proved as one of the most viable semiconductor material systems for tunable optoelectronics which is expected to bring forth a similar impact at nanoscale. Thin layers and multilayer heterostructures of group III nitrides in a wurtzite structure have been intensively developed using metalorganic chemical vapor deposition (MOCVD) processes leading to "the invention of efficient blue light-emitting diodes which has enabled bright and energy-saving white light sources" (The Nobel Prize in Physics 2014) [1]. It has been predicted by first-principles calculations that the band gap of 2D group III nitrides in a honeycomb monolayer structure across the compositional range of BN, AlN, GaN, InN, and TlN can be tuned from UV to IR, and even further to the THz frequency range [2, 3]. 2D InGaN and InTlN have further been suggested for efficient light harvesting thus boosting opportunities for photovoltaic applications [2]. It is to be noted that InN, in its 2D $sp^2$-bonded honeycomb monolayer structure, and also in its $sp^3$-bonded wurtzite bulk crystal, exhibits the smallest effective electron mass among the group III nitrides [2]. In general, small effective electron mass relates to high electron mobility and saturation drift velocity which are properties of crucial importance for enabling evermore high-speed and high-frequency performance of (nano)electronics. Similar prospects are being held by 2D InO [4, 5] which - among a group of several monolayer metal oxides of various atomic structures - has also been predicted by first-principles calculations to exhibit ultrahigh electron mobility [6].

Noteworthy, by now both the 2D InN [7] and the 2D InO [4] have crossed over the realm of predictive first-principles calculations to their material realization in confinement at graphene/SiC interface via MOCVD.

As already pointed out, MOCVD has been established as a prime deposition technology for obtaining (ultra)thin films and heterostructures of semiconductor materials such as group III nitrides [1] and group III arsenides [8], which build up components as the semiconductor heterostructure lasers and transistors giving the access of our modern high-tech society to fiber-optic and satellite communications [8]. Accordingly, an advance of 2D materials and device heterostructures for perceived electronic applications, which can make an impact on our ever day life, depends on developing their growth by MOCVD. Thus far, examples of MOCVD of 2D materials have been scattered.

In what concerns group III nitrides, the conceptually novel 2D InN [7], GaN [9], and AlN [10] have been demonstrated by MOCVD processes assisted by the employment of SiC substrate covered by graphene obtained in a high-temperature sublimation process. Intercalation of atoms at the graphene/SiC interface has been acknowledged *sine qua non* for the accomplishment of 2D group III nitrides by MOCVD. A defining characteristic of MOCVD is the employment of metalorganic precursors such as trimethyl-indium, -gallium, and -aluminum, which contain (strong) metal-carbon bonds. Therefore, it is of crucial importance to locate the mechanisms responsible for metalorganic precursor/graphene reactions which can deliver group-III adatoms or group-III-containing admolecules on graphene from common gas-phase methyl-based precursors. Atomistic pathways that control MOCVD processes are largely unknown.

In this work, we use ab initio molecular dynamics (AIMD) accounting for van der Waals corrections to simulate trimethylindium $(CH_3)_3In$ (or TMIn)/graphene reactions and to gain knowledge about reaction dynamics for supply of indium (In) atoms to the graphene surface. We extend a model which incorporates SiC-supported graphene which is of relevance for understanding the reactions of MOCVD precursors on regions of the zero (buffer-layer) graphene directly exposed to the gas phase. We explore, in addition, a comparison with our recent *ab initio* molecular dynamics simulations which has revealed atomistic and electronic mechanisms that govern surface reactions of trimethylaluminum $(CH_3)_3Al$ (or TMAl) with defect-free self-standing graphene and produce isolated Al adatoms [11]. We note that other, classical molecular dynamics simulations supported by experiments indicate plausible atomistic pathways for Ga atoms intercalation across extended defects in graphene [12].

## 2. METHODS AND COMPUTATIONAL DETAILS

Born-Oppenheimer *ab initio* molecular dynamics simulations are carried out with the VASP code [13] using the local density approximation (LDA) [14] and the projector augmented wave [15] method. The approximation proposed by Grimme [16] is employed to describe van der Waals interactions. At each AIMD time step (0.1 fs), the total energy is evaluated to an accuracy of $10^{-5}$ eV/supercell using Γ-point sampling of the reciprocal space and a planewave energy cutoff of 300 eV. $k_BT$ Gaussian smearing of electronic states is used to mimic the electronic temperature.

We have performed a set of investigations of TMIn + 9 $H_2$ gas molecules impinging on a graphene/SiC substrate at 3500 K. The use of high temperatures is justified by the relatively low chemical reactivity of TMIn and by the fact that Born-Oppenheimer molecular dynamics is highly computationally intensive. However, structural stability of graphene at these temperatures is ensured throughout our AIMD runs [see **Figure 1S** in the *Supplementary Material*]. In our AIMD supercell model, the SiC substrate is passivated underneath by static H atoms and bonded on top to a zero-layer graphene – known also as a buffer layer or a interfacial layer – which is a result to the formation mechanism of graphene by thermal decomposition of SiC [7]. The initial surface slab $4\sqrt{3} \times 4\sqrt{3}$ R30° graphene/SiC, illustrated in **Figure 1**, is obtained by full structural relaxation at 0 K via density-functional theory (DFT) using total-energy and force convergence criteria of $10^{-5}$ eV/supercell and 0.01 eV $Å^{-1}$, respectively. The graphene/SiC slab is formed of 246 C, 96 Si, and 48 H atoms and has lateral sizes of 2.47 nm. Although our SiC slab is not sufficiently thick to distinguish between hexagonal or cubic stacking sequences, we will refer to this layer as SiC (0001), which is indicative of a typical surface termination of hexagonal SiC used in MOCVD experiments. Our simulation box (graphene/SiC substrate + gas molecules (9 $H_2$ and $(CH_3)_3In$) contains 421 atoms in total.

The methods used to simulate TMAl reactions on defect-free graphene are detailed in Ref. [11]. An analogous procedure is employed here to simulate TMIn precursor reactions on free-standing graphene as well as on graphene/SiC substrates. First, the geometries of chemical precursors are optimized via 0-kelvin DFT energy minimization. Assigned random velocities to each atom (translational kinetic energy corresponding to initial temperature of 300 K), the dynamics of the precursors is followed for ≈2 ps using microcanonical NVE sampling and 0.1 fs timesteps. In parallel (separate simulations), the graphene/SiC substrate is equilibrated at 3500 K using NVT sampling (Nose-Hoover thermostat), while keeping static H atoms that passivate the bottom SiC layer. Hence, the precursor internal atomic positions and velocities, as well as the positions and velocities of the graphene/SiC slab obtained from initial AIMD runs are used as input for AIMD simulations of MOCVD processes.

**Figure 1**. Orthographic plan view (a) and cross-section views (b) and (c) of zero-layer graphene/SiC (0001) surface slab after DFT relaxation at 0 K. The visualization of chemical bonds applies cutoff lengths of 2.2 Å. The crystallographic directions refer to graphene (graphite) lattice axes.

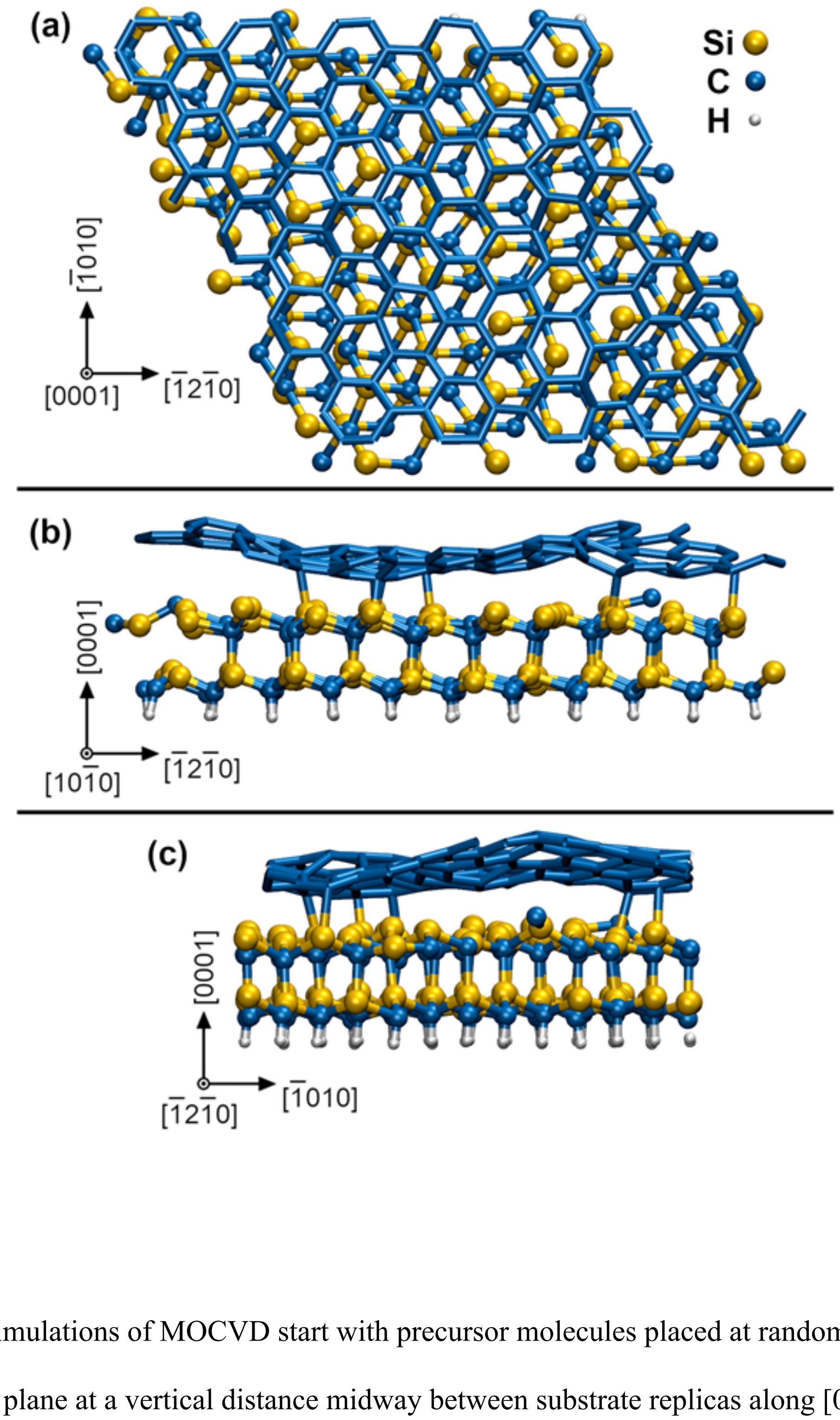

AIMD simulations of MOCVD start with precursor molecules placed at random positions on a (0001) plane at a vertical distance midway between substrate replicas along [0001] (see crystallographic directions in **Figure 1**). Given that input-velocities are taken from NVE simulations of molecules in vacuum, the gas precursors initially have zero mass-centrum drift. Therefore, during the initial 10 fs of AIMD/MOCVD simulations, the precursors are accelerated toward the surface using constant gentle forces ($10^{-2}$, $10^{-3}$ and $10^{-4}$ eV/Å for In, C and H atoms, respectively). This feature is implemented an in-house modified VASP code [17]. Then, the forces are removed, and the dynamics of the gas molecules is integrated on an NVE scheme, while the substrate is subject to thermostatting at 3500 K. The possibility of combining NVE (for precursors) and NVT (for graphene/SiC) sampling within the same AIMD simulation is implemented in our modified version of VASP [17]. Directly coupling the internal degrees of freedom of the molecules with a thermostat would alter their dynamics, thus causing artifacts such as sudden molecule splitting (as, e.g., relatively facile $N_2$ dissociation in the gas phase [18]). Note that, during our AIMD simulations, the molecules reach thermal equilibrium via collisions with the substrate.

AIMD snapshots and figures are generated using the VMD software [19]. In figures, “dynamic bonds” have typical cutoff lengths of 2.2 Å. Note, however, that some bonds are manually removed to facilitate visualization of molecular structures. We caution the reader that our descriptions of molecular reactions are based on own interpretations of electron transfer processes: it is known that density-functional theory is inherently limited (treats the electron density, not individual electrons) in the description of dissociation of asymmetric molecules [20].

## 3. RESULTS

Our investigation focuses on a model which incorporates the zero-layer graphene exposed to TMIn and hydrogen molecules. This is the interfacial (buffer) layer between SiC (0001) substrate and the 1st graphene layer, and it is a result to the mechanism of formation of graphene by thermal decomposition of SiC [7]. Therefore, besides allowing to understand precursor reactions responsible for nucleation phenomena, the results of our simulations are also starting point for dedicated atomic-scale investigation of intercalation phenomena. In this regard, we note that the zero-layer graphene is directly exposed to the gas near SiC surface steps (see illustration in figure 6 in Ref. 21) or at local terminations of the top graphene layer (schematic illustration in figures 6, 7, 8 in Ref. 22). Adatoms or admolecules – produced via MOCVD on the zero-layer graphene – may find intercalation pathways at defective sites of the zero-graphene layer. Conversely, the reactivity of the top ($1^{st}$) graphene layer can be well described by a free-standing graphene sheet (the interactions of graphene with an underlying graphene layer is weak). The results of TMIn reactions on self-standing defect-free graphene can be found in the *Supplementary Material*. In general, it is expected that zero-layer graphene is more reactive than self-standing graphene due to bonding with underlying Si atoms of the SiC substrate.

We carry out 7 independent simulations at 3500 K, for a total simulated time of 77 ps (770 000 configurations considering a time step = 0.1 fs). Each simulation has approximately the same duration (≈11 ps). In 3 of out 7 simulations (which we name **Simulation #5, #6, and #7**), the TMIn precursor does not dissociate for the entire duration of the AIMD run. During those simulations – which correspond to a total simulated time of approximately 30 ps – we record 16 TMIn/graphene adsorption and desorption events. Isolated Indium adatoms and/or methyl-In admolecules form in 4 out of 7 cases (**Simulation #1, #2, #3, and #4**). More in general, our AIMD simulations reveal that TMIn + $H_2$ + graphene reactions lead to formation of products including indium hydride (InH), methane, ethane, ethene, and propane.

**Simulation#1.** In the initial 8.2 ps of this simulation, the trimethylindium $(CH_3)_3In$ (TMIn), precursor impinges on graphene on five occasions without reacting. During this time, however, the TMIn molecule remains near (bounces on) the graphene surface. This is probably due to the unoccupied electrophile $sp^3$ Indium orbital, which temporarily accepts electrons from graphene $\pi$ states or unsaturated carbon $sp^3$ orbitals (note that some graphene C atoms are bonded to Si). TMIn/graphene interactions – either of van der Waals type or direct molecule/surface collisions – excite vibrational states of the molecule which ultimately activate the first TMIn reaction. The sequence of events illustrated in **Figure 2** occurs during a timeframe of ≈5 ps. The reactions start with $H^+$ detachment from a methyl group, **Figure 2(a)**, which produces a $(CH_3)_2In^-{=}CH_2$ negative gas ion and protonation of graphene, **Figure 2(b)**. Then, a methyl group detaches from the central Indium atom to attach to the $CH_2$ group, **Figure 2(c)**. The reaction, interpreted in **Figure 3(a, b)**, is the result of various electron-transfer events. A pair of electrons – one transferred from the $\pi$ $In(p_z)$–$CH_2(p_z)$ HOMO state and the other from the $\sigma$ $In(sp^2)$–$Methyl(sp^3)$ state – forms a $CH_2(sp^3)$–$Methyl(sp^3)$ bond, thus producing a $H_3C$–$In^-$–$CH_2$–$CH_3$ ion. The intramolecular transfer of a methyl group leaves two free electrons on Indium. In principle, each of these electrons may occupy an individual non-bonding $sp^3$ orbital to minimize Coulombic repulsion. However, due to vicinity of a proton (adsorbed on graphene) the electrons pair up in one of the $sp^3$ dangling orbitals, as shown in **Figure 2(b)**. The free electron pair attracts the proton, which thus desorbs from the graphene surface and bonds with the Indium atom, **Figure 2(c)**, and **Figure 3(c, d)**. The process leads to formation of a $H_3C$–InH–$CH_2$–$CH_3$ neutral molecule, with all orbitals hybridized $sp^3$, **Figure 3(d)**. Soon after (<1 ps), however, the $H_3C$–InH–$CH_2$–$CH_3$ species dissociates into $H_3C\dot{I}nH$ and $\dot{C}H_2CH_3$ radical molecules, see **Figure 3(e, f)** and **Figure 2(d)**. After temporarily (≈0.5 ps) bonding to graphene, **Figure 2(e)**, the $\dot{C}H_2CH_3$ radical attacks the methyl group of $H_3C\dot{I}nH$, **Figure 3(g)**. The reaction leads to formation of indium hydride (InH) and propane ($C_3H_8$) molecule**, Figure 3(h)**, which float away from the surface, **Figure 2(f, g)**.

**Figure 2**. (**Simulation#1**) Sequence of trimethylindium $(CH_3)_3In$ reactions on graphene/SiC (0001) as observed in AIMD simulations at 3500 K. The dynamic bonds have cutoff lengths of 2.2 Å.

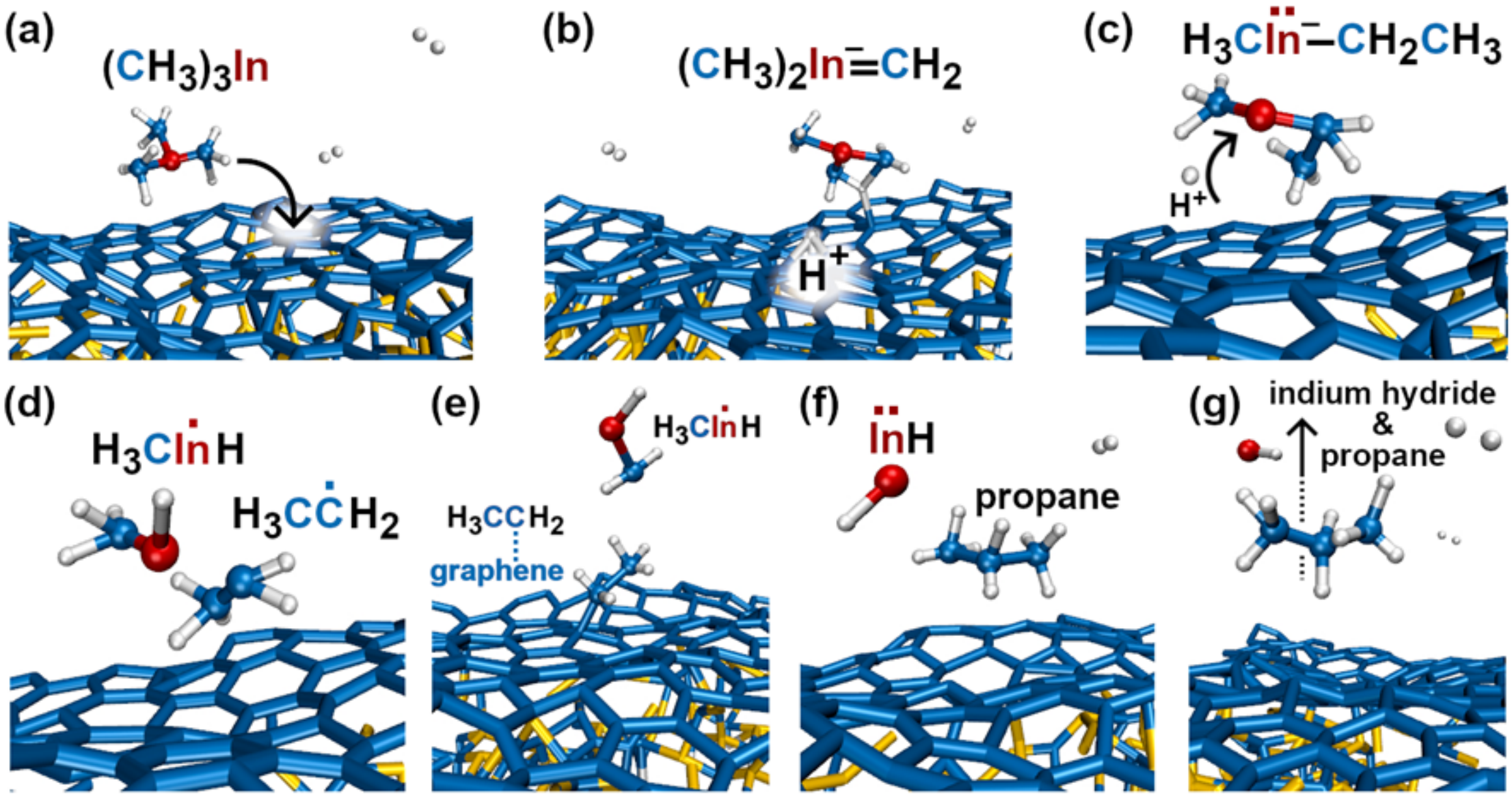

**Figure 3**. (**Simulation#1**) Trimethylindium $(CH_3)_3In$/graphene reactions lead to formation of indium hydride and propane gas molecules.

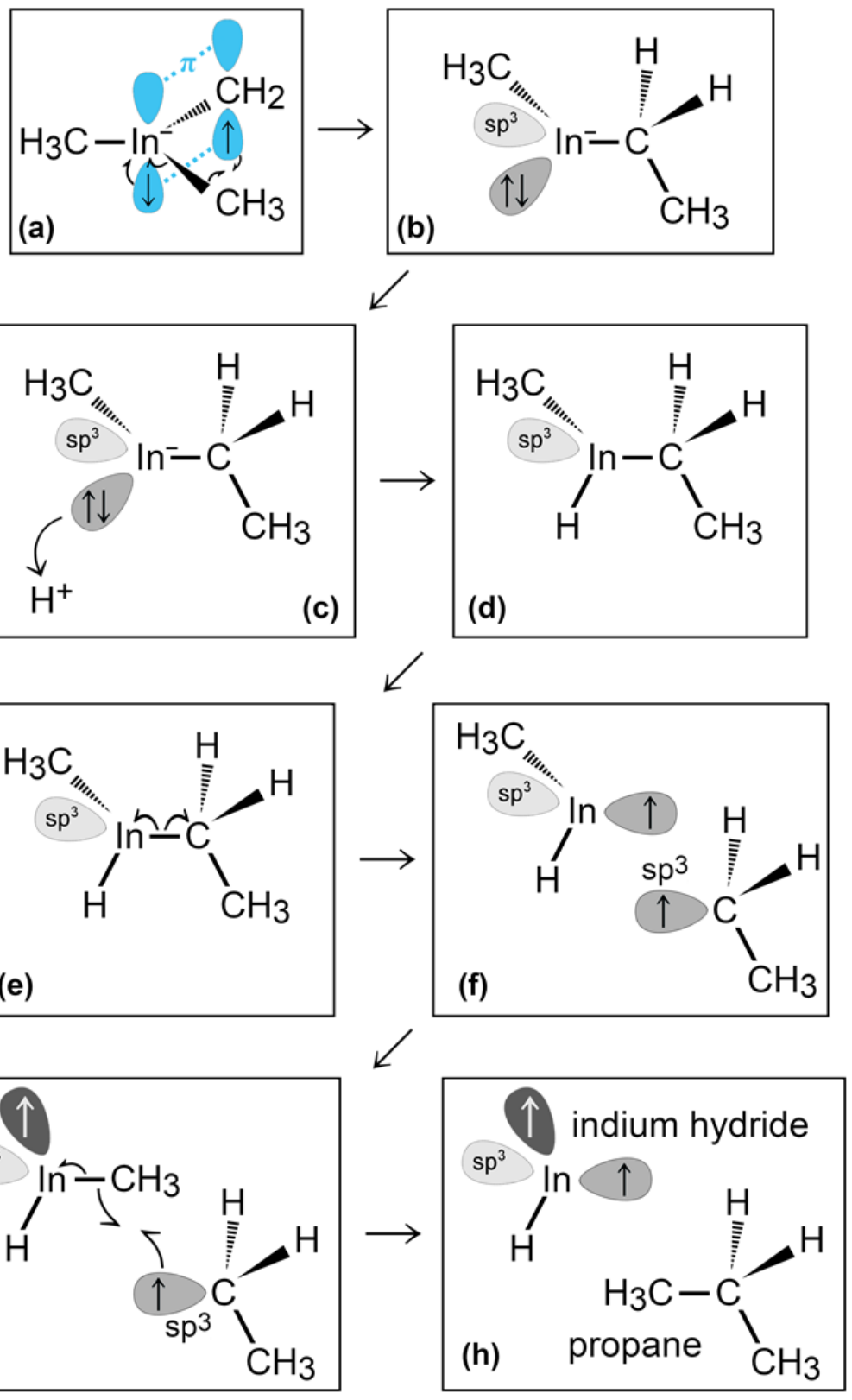

To gain insights on the reactivity of an InH gas molecule with graphene, **Figure 4(a)**, the simulation is extended for ≈4.5 additional ps. After two InH molecule bounces, the proton of the InH molecule is attracted by the π electron cloud of graphene, **Figure 4(b)**, and detaches from the In ion ($In^-$). Both the proton and $In^-$ ion remain on graphene as adspecies, **Figure 4(c)**. After a fraction of ps, the propane gas molecule (previously formed), see **Figure 2(f, g)**, approaches the $In^-$ adspecies and donates a proton from one of its terminal methyl groups. The reaction produces an InH admolecule, **Figure 4(d, e)**, while the electron pair formerly occupying the C-H bond transfers to a non-bonding $sp^3$ carbon orbital of the $CH_3CH_2\ddot{C}^-H_2$ molecule, **Figure 4(e)**. Then, within one ps, the $sp^3$ electron pair of $\ddot{C}^-$ takes a proton from the central $CH_2$ group, **Figure 4(f)**, thus producing $CH_3\ddot{C}^-HCH_3$, **Figure 4(g)**. Accidentally, the $CH_3\ddot{C}^-HCH_3$ species flies near the InH admolecule. The H atom of the central C-H bond detaches from the molecule and bonds to the proton of InH, **Figure 4(g)**. The reaction leads to elimination of a $H_2$ gas molecule and reforms an $In^-$ monomer, **Figure 4(h)**. The process also produces a highly reactive $CH_3$-$\ddot{C}$-$CH_3$ molecule, which quickly transforms in a $\dot{C}H_2CH_2\dot{C}H_2$ double radical via intramolecular hydrogen transfer, **Figure 4(i)**.

**Figure 4.** (**Simulation#1**) Indium hydride InH and propane reactions on graphene/SiC (0001) as observed in AIMD simulations at 3500 K. The reactions end with formation of an $In^-$ adspecies. The dynamic bonds have cutoff lengths of 2.2 Å. The distance of the $In^-$ monomer from the closest C atoms of graphene is typically within the range 2.3 – 2.5 Å. The entire sequence of events – from InH arrival on graphene (a) to formation of $In^-$ and $H^+$ adspecies, $H_2$ and $C_3H_6$ gas molecules (i) – occurs over a timeframe of ≈2 ps.

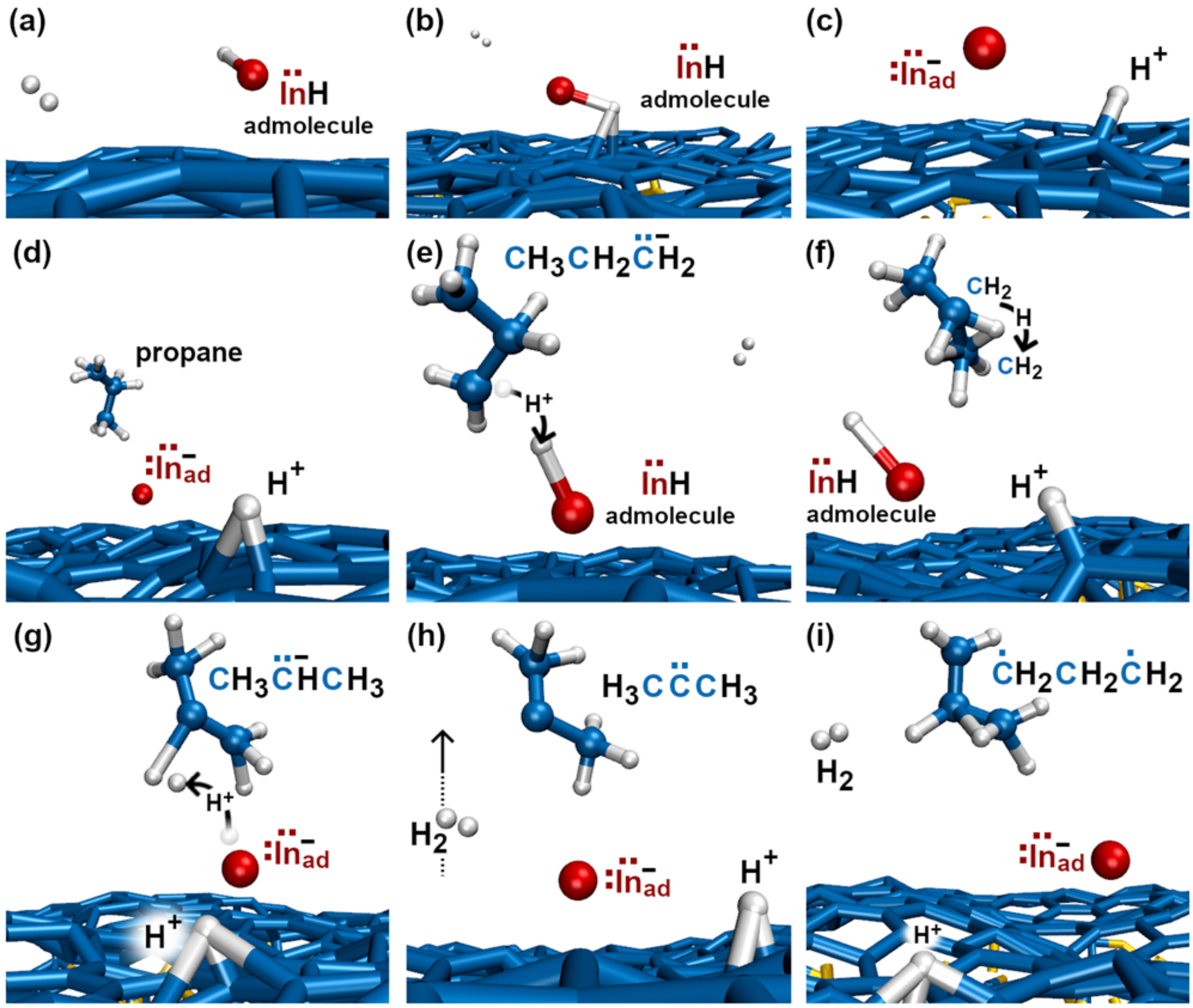

**Simulation#2.** In this simulation we observe three collisions of TMIn onto graphene during 8.5 ps. The remaining 3.5 ps of this simulation reveal several intricate reaction pathways which ultimately lead to formation of an $In^{-}$ monomer or an InH admolecule on graphene.

The first relevant reaction event of **Simulation#2** starts with a hydrogen molecule hitting a methyl group of TMIn, **Figure 5(a, b)**. The momentum transfer causes bending of methyl-In-methyl chemical bonds, thus leading one methyl group to rapidly approach, and to react with, another methyl, **Figure 5(c)**. The process produces an ethane molecule ($C_2H_6$) and a methyl-In molecule, **Figure 5(d, e)**. Between **Figure 5** and **Figure 6**, we observe several intra- and inter-molecular proton transfer events which are mediated by Indium. Starting, for example, from the configuration shown in **Figure 5(e)**, the Indium atom of the $CH_3In$ molecule takes a proton (to later give it back) from the underlying ethane molecule. Due to these frequent proton exchanges, the ethane molecule transforms back and forth into a $:C^{-}H_2CH_3$ species, which is relatively strongly bonded to graphene.

Then, we observe a rapid proton exchange between the two molecules: an $H^+$ leaves a methyl group of ethane to attach to the In atom, thus temporarily forming a $CH_3In^+H$ molecule (figure not shown). After 0.2 ps, the proton returns to its former position, assisting detachment of the ethane molecule from graphene, **Figure 5(e), and Figure 6(a)**. The subsequent approach of $C_2H_6$ to $InCH_3$ induces intramolecular proton transfer within the ethane molecule, $H^+$ jump indicated by arrow in **Figure 6(b)**. The reaction is interpreted as described by **Figure 7**. The electron pair on In overlaps with the antibonding σ* $sp^3$(C)–$sp^3$(C) molecular orbital of ethane, represented by orange lobes in **Figure 7(a)**. The electron occupation of the σ* state weakens the C-C bond and assists proton transfer from one carbon to the other, **Figure 7(b)**. The reaction ultimately leads to formation of methane and $H_2C{=}InCH_3$ molecules, **Figure 7(c, d)**, which rapidly leave graphene surface, **Figure 6(c, d)**.

**Figure 5. (Simulation#2)** $H_2$-methyl collision-activated TMIn dissociation to ethane and methyl-indium.

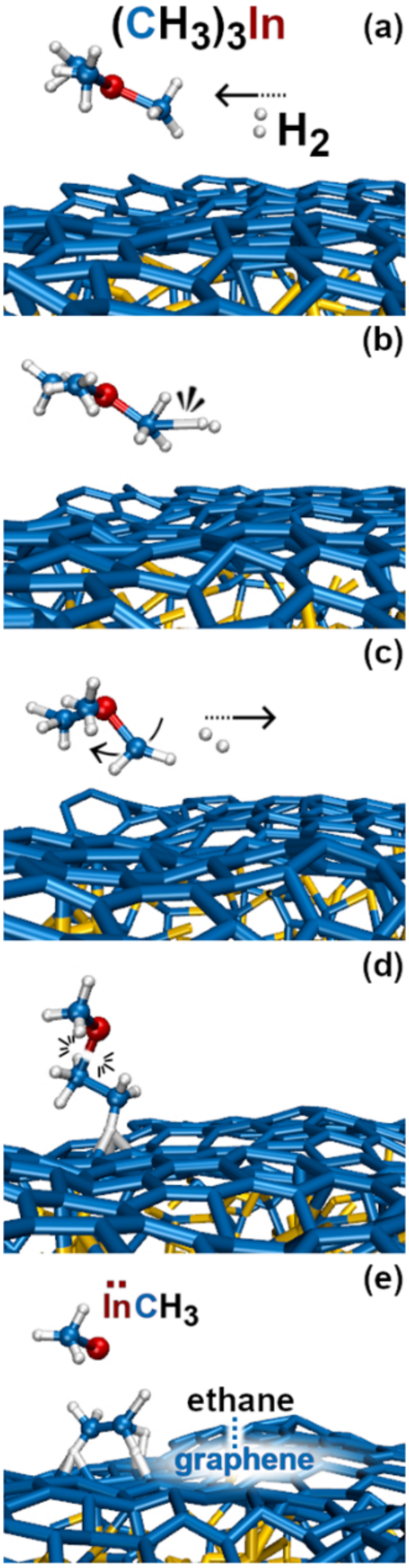

**Figure 6. (Simulation#2)** Methane formation via methyl-In-assisted ethane dissociation.

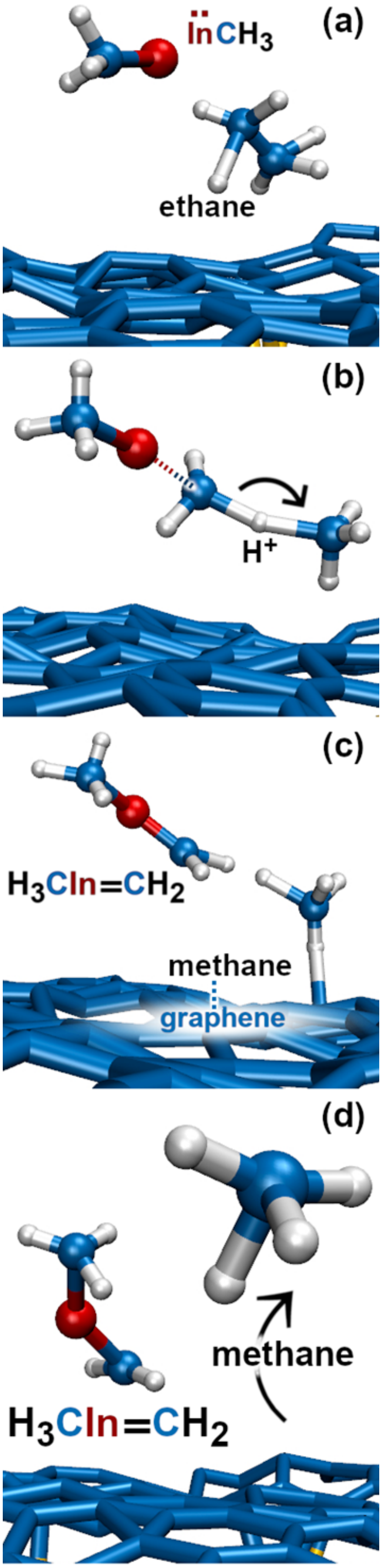

**Figure 7. (Simulation#2)** Formation of methane and $H_2C{=}InCH_3$ due to methyl-In reaction with ethane.

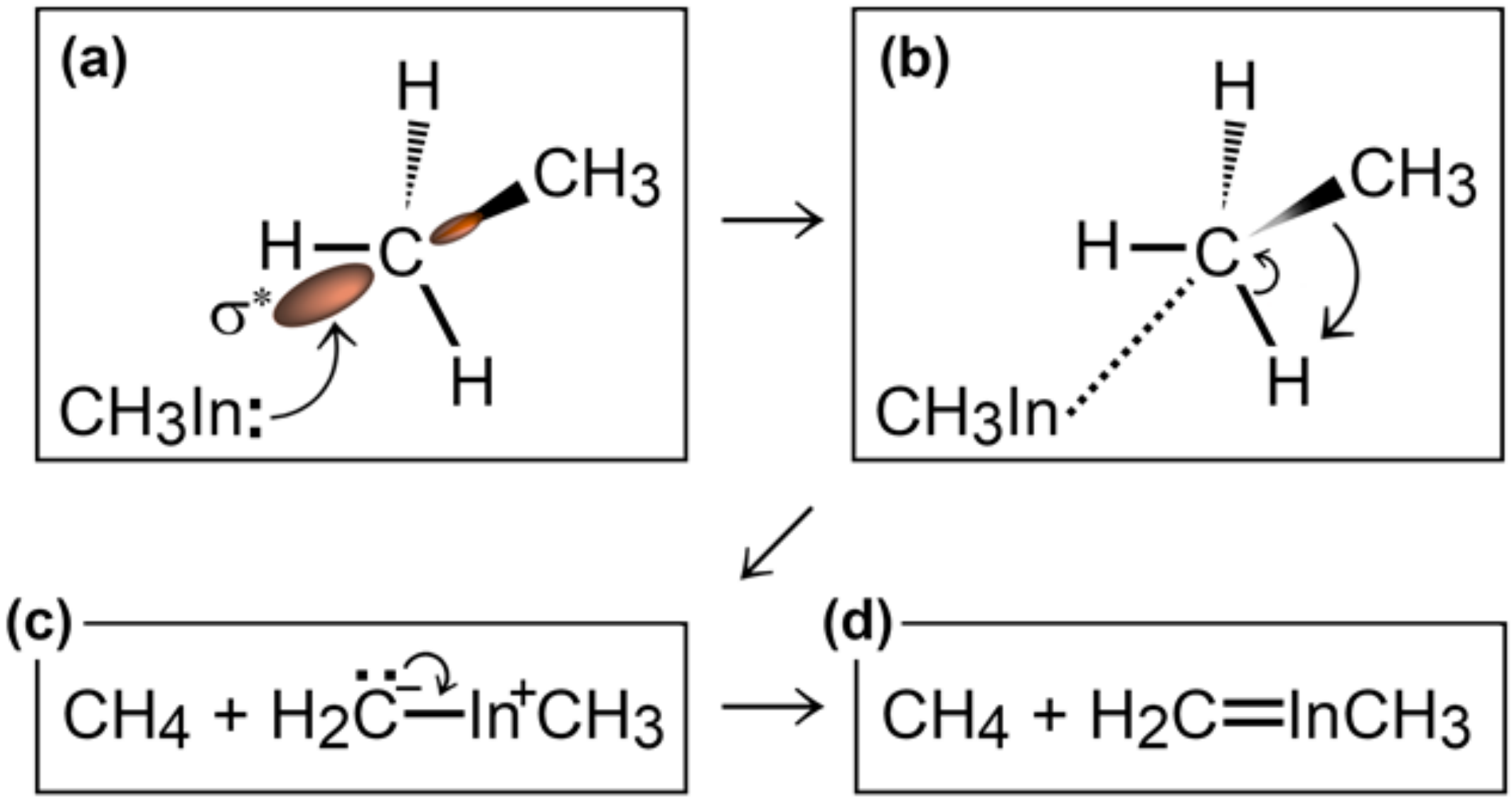

During the subsequent 0.5 ps, the $H_2C{=}InCH_3$ molecule undergoes several transformations, as illustrated in **Figure 8**. Within 0.1 ps, we observe two back-and-forth proton exchanges from the methyl group to the In atom, temporarily forming a $\dot{C}H_2InH\dot{C}H_2$ double radical (not shown). After ≈0.2 ps, however, a proton permanently leaves the methyl group, **Figure 8(a, b)**, thus producing $\ddot{C}H_2$ and a $H_2C{=}InH$ molecules, **Figure 8(c)**. The highly reactive $\ddot{C}H_2$ molecule quickly attacks a graphene carbon atom, indicated as $C_G$ in **Figure 8(d)**, to form a $C–C_G$ bond. We attribute formation of this bond as due to both C and $C_G$ atoms contributing one electron each, which produces a free radical electron on the admolecule. The dangling $CH_2$ group is prone to react with the near $H_2C{=}InH$ gas molecule. Indeed, the subsequent ≈0.1 ps of the simulation shows that: first, the $H_2C{=}InH$ molecule bonds to the $CH_2$ group attached on graphene, **Figure 8(e)**; then, the two $CH_2$ groups form an ethene molecule which rapidly desorbs, leaving an InH admolecule on graphene, **Figure 8(f, g)**. The atomistic pathways shown in **Figure 8(e, f, g)** are rationalized by the reaction diagram of **Figure 9**. The $\dot{C}$ atom of the $\dot{C}H_2$ radical attached on graphene attacks the π orbital of the $H_2C$=InH molecule, **Figure 9(a)**. The transition state depicted in **Figure 9(b)** evolves into HC=CH and InH products, **Figure 9(c)**. However, while ethene leaves the surface, the InH molecule remains adsorbed on graphene for the remaining part (≈2 ps) of the simulation. During this timeframe, we repeatedly (three events) observe InH admolecule splitting followed by $In^-$ and $H^+$ adspecies recombination, **Figure 8(h, i)**. Likely due to electrostatic attraction, $In^-$ and $H^+$ adspecies remain within few Å from each other.

**Figure 8. (Simulation#2)** AIMD snapshots of reactions leading to formation of InH and $In^-$ adspecies.

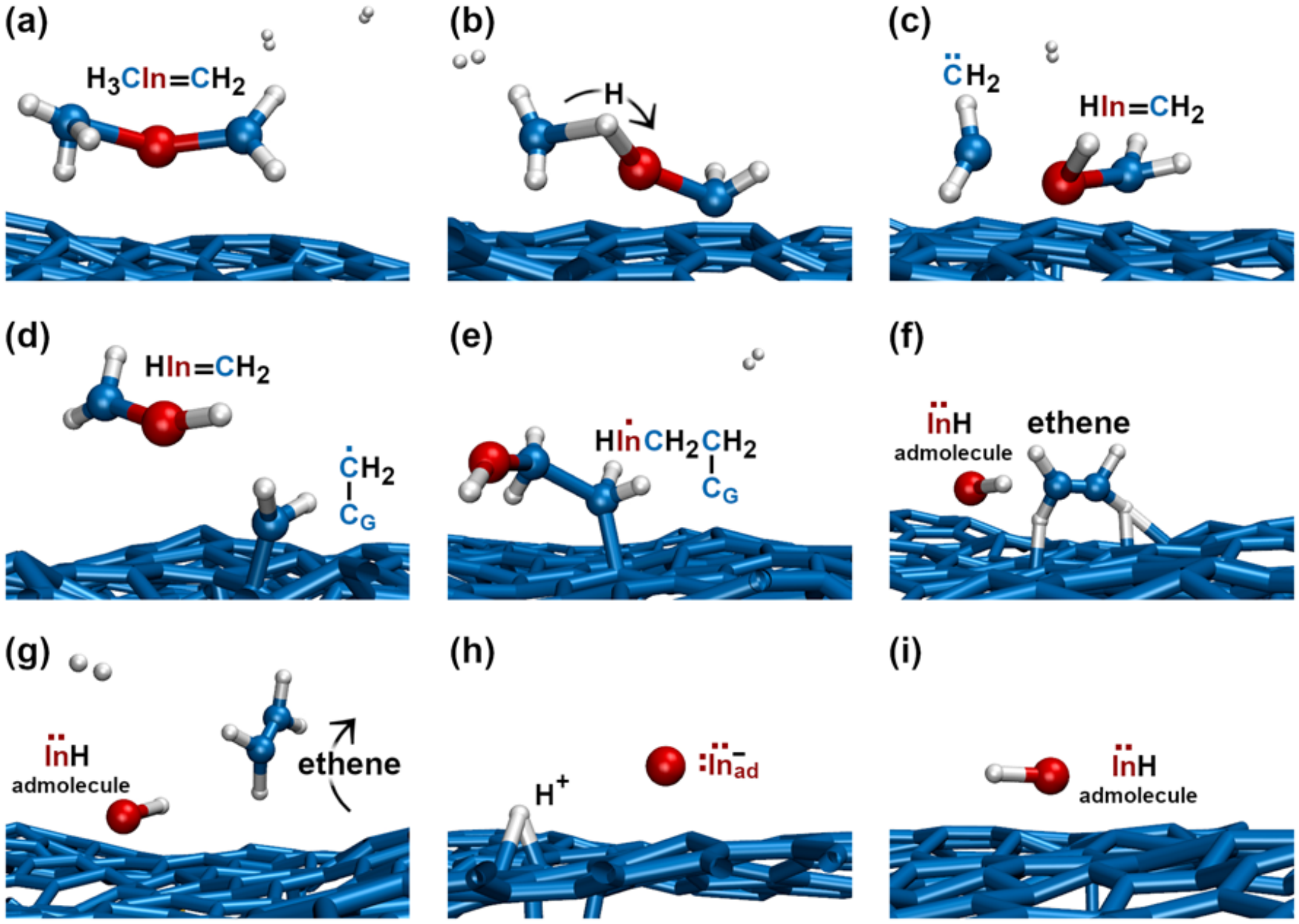

**Figure 9. (Simulation#2)** Interpretation of reactions between a $\dot{C}H_2$ radical adsorbed on graphene and a $H_2C{=}InH$ gas molecule leading to formation of InH.

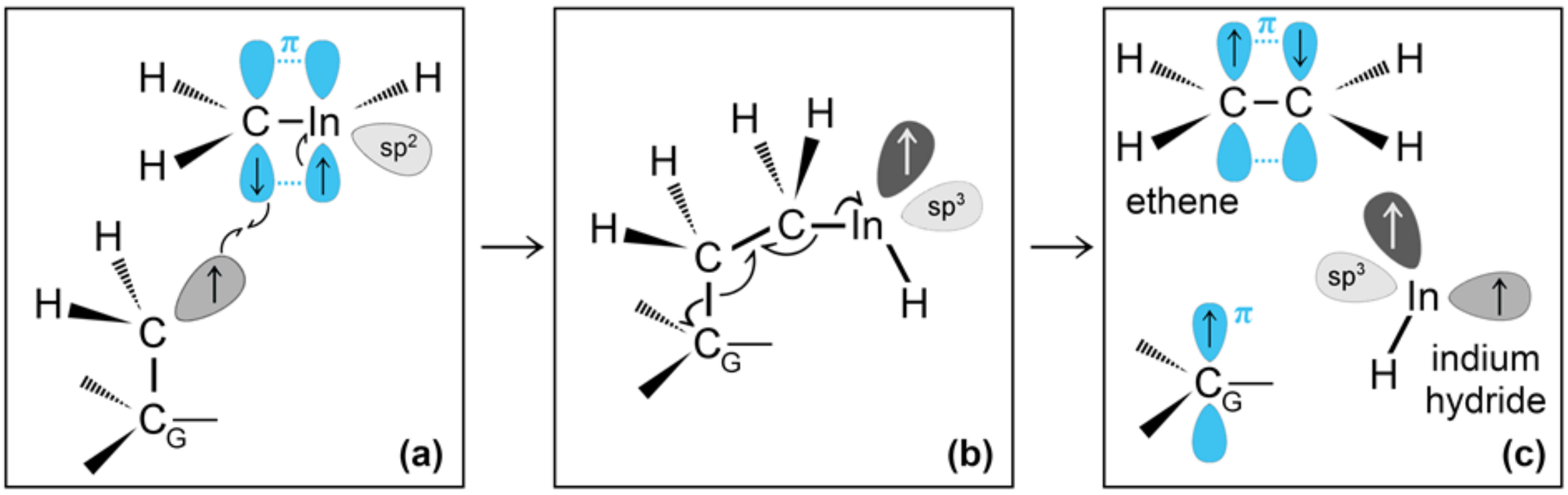

**Simulation#3.** The first relevant TMIn reaction occurs in the gas phase, at a simulation time of ≈3 ps, when a $H_2$ molecule hits TMIn, **Figure 10(b)**. The impact assists detachment of a methyl group, **Figure 10(c)**, which adsorbs on graphene for ≈1 ps, while the $(CH_3)_2In$ molecule remains in the gas, **Figure 10(d)**. At a simulation time of ≈4.1 ps, the methyl radical desorbs from graphene, **Figure 10(e)**, attaches to graphene and further reacts with a hydrogen molecule, **Figure 10(f, g)**, thus producing methane ($CH_4$) and atomic hydrogen, **Figure 10(h)**. Nearly at the same time, TMIn adsorbs on graphene, **Figure 10(i)**. The interaction with the surface excites vibrational modes which promote detachment of a methyl group, **Figure 10(j)**. Both the methyl and $CH_3In$ molecules remain adsorbed and migrate on graphene. However, after 0.7 ps, the methyl radical flies away from the surface, **Figure 10(k)**. Finally, the $CH_3In$ admolecule reacts further by eliminating the remaining methyl group. The latter reaction leaves a neutral In adatom which migrates on graphene during the remaining simulation time of ≈2 ps, **Figure 10(l)**. Conversely, the $\dot{C}H_3$ adspecies diffuses on the surface during 0.4 ps and then desorbs, **Figure 10(l)**.

**Figure 10. (Simulation#3)** Sequential elimination of methyl groups from TMIn leads to formation of an electrically neutral In adatom on graphene.

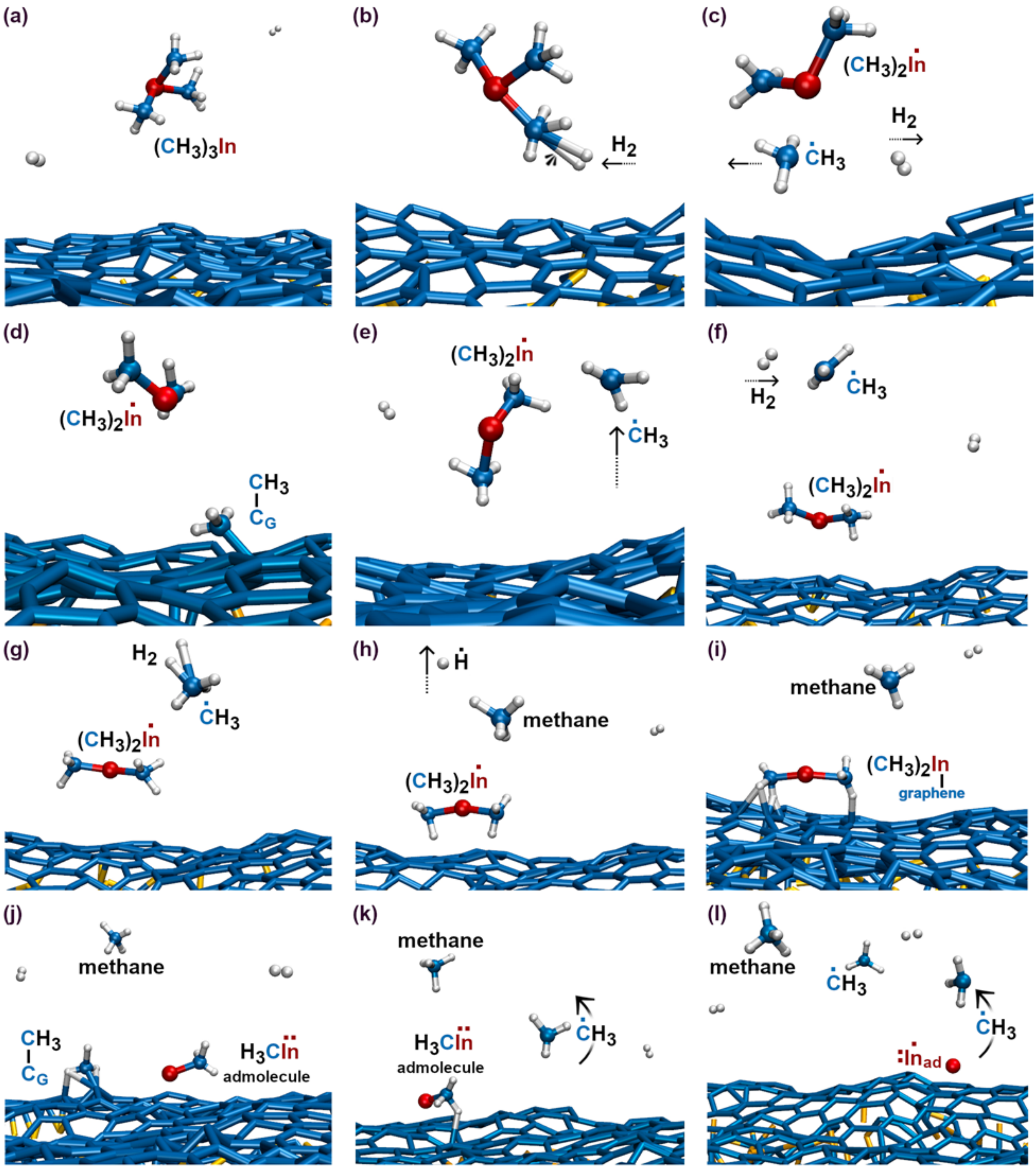

**Simulation#4.** The first TMIn/graphene collision, **Figure 11(a, b)** – which occurs ≈1 ps after thermal equilibration of the gas/substrate system – leads directly to elimination of a methyl radical from TMIn, **Figure 11(c)**. The $In(CH_3)_2$ molecule remains adsorbed on graphene,

**Figure 11(d)**. However, after ≈1.5 ps, the interaction with graphene's surface causes elimination of an additional $\dot{C}H_3$ group, **Figure 11(d, e)**. The methyl molecule attaches and diffuses on graphene during ≈2.3 ps. Then it desorbs from the surface, **Figure 11(f)**. It is worth noting that the methyl gas molecule exchanges a proton with a $H_2$ molecule which flies nearby (figure not shown). The exchange reaction has as intermediate transition state a molecular complex formed of methane and atomic hydrogen. During the remaining 9.2 ps of **Simulation#4**, the methyl-Indium admolecule bounces on the surface 5 times without dissociating.

**Figure 11. (Simulation#4)** After 4.1 ps all reactions have occurred the methyl-In remains as admolecule for the remaining 9.2 ps of the simulation.

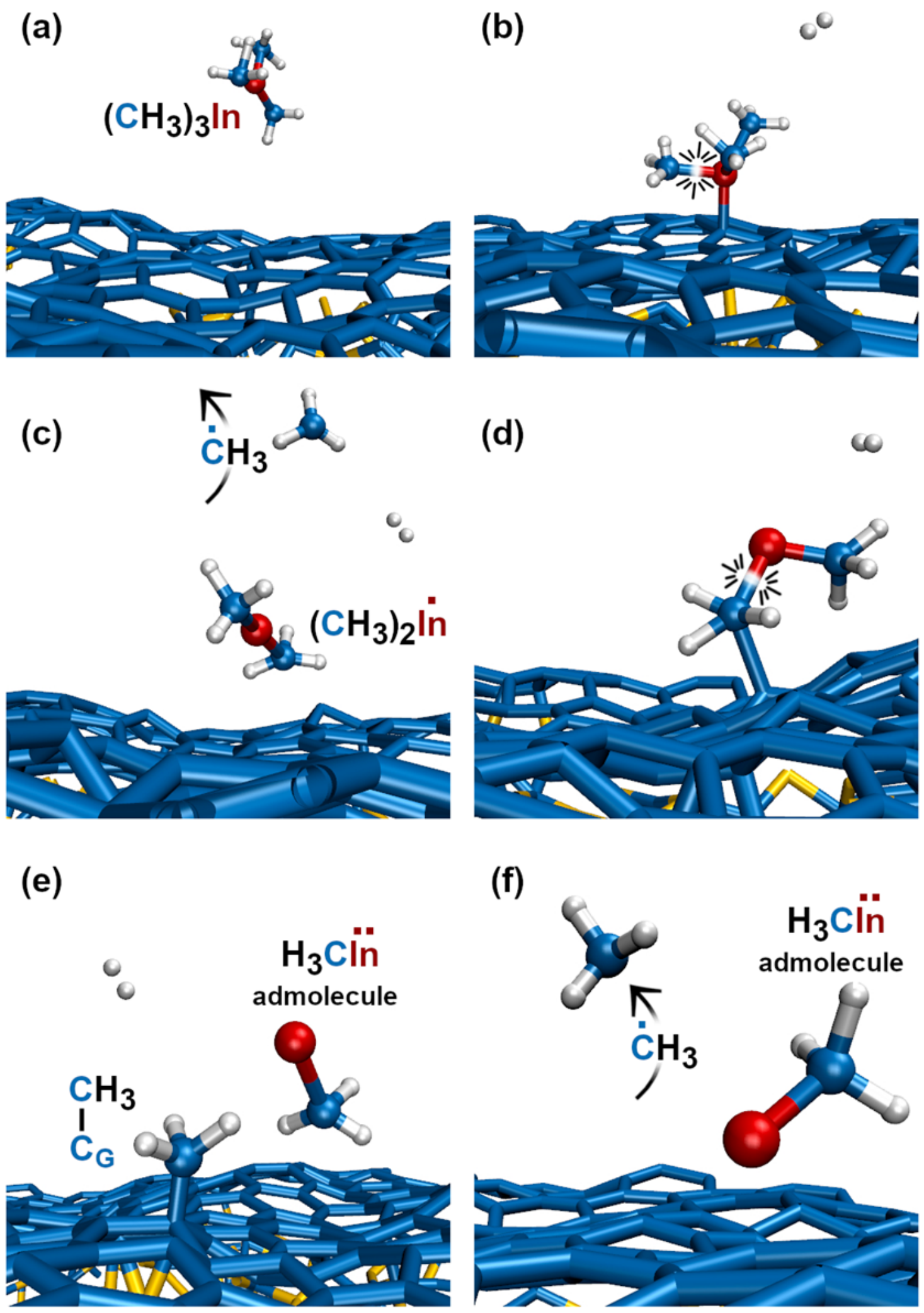

## 4. DISCUSSION

To summarize, relevant intermediate and final products of TMIn+$H_2$ reactions on graphene/SiC (0001) in **Simulation#1** are: InH gas molecules, **Figure 2(f, g)** and admolecules, **Figure 4(a, f)**, charged $In^-$ and $H^+$ adatoms, **Figure 4(c, d, h)**, $H_2$, **Figure 4(h)**, and propane, **Figure 2(f, g)**, gas molecules. **Simulation#2** also yields InH admolecules – that dissociate (and recombine) in $In^- + H^+$ adspecies, **Figure 8(h, i)** – as products of TMIn/graphene/SiC(0001) reactions. However, at variance with **Simulation#1**, the first reaction of **Simulation#2** is triggered by a collision between a $H_2$ molecule and TMIn in the gas phase (**Figure 5**a,b). In addition, **Simulation#2** evidences formation of ethane, **Figure 6(a)**, methane, **Figure 6(d)**, and ethene, **Figure 8(g)**, gas molecules. **Simulation#3** shows formation of electrically neutral In adatoms on graphene, **Figure 10(l)**. In this case, formation of an In adatom stems from sequential elimination of methyl-group radicals from the TMIn molecule. Analogous to **Simulation#2**, transformation of TMIn initiates (first detachment of a $\dot{C}H_3$ group) in the gas phase due to collision with a $H_2$ molecule, **Figure 10(b, c)**. The graphene layer catalyzes elimination of the other two $\dot{C}H_3$-groups, **Figure 10(i–l)**. In addition, it is worth underlining a gas-phase reaction between a $\dot{C}H_3$ radical (formerly part of the TMIn precursor) and a $H_2$ molecule which produces methane and atomic hydrogen, **Figure 10(g, h)**. In **Simulation#4**, TMIn reacts only on the graphene surface. The simulation shows two relatively rapid (within 4.1 ps) eliminations of $\dot{C}H_3$ groups, **Figure 11**. In this case, however, the remaining methyl-In admolecule diffuses on the graphene surface without reacting for the remaining part of the simulation (≈9.2 ps).

Since the MOCVD process involves the trimethylindium precursor in presence of molecular hydrogen, both species are considered in this work for achieving realistic descriptions of reactions in the gas phase and on surface. At variance with what observed during TMAl [11] and TMIn reactions on self-standing defect-free graphene (see *Supplementary Material*), our present simulation model that includes $H_2$ gas molecules allows us to gain insights into the effects of hydrogen in MOCVD processes of TMIn on graphene. AIMD simulations show that TMIn reactions are mediated not only by the graphene surface: in some occasions, TMIn reactions are triggered in the gas phase by the energy transfer provided through collision with a hydrogen molecule, **Figures 5(b-d)** and **10(b, c)**. These events underline the role of hydrogen in providing the energy needed for precursor molecules to enter in a transition state. The contribution of hydrogen to the reactivity of TMIn on graphene is also indicated by transfer of a proton (previously adsorbed on graphene) to In-containing molecules that fly proximate to the surface, see, e.g, **Figure 2(b-d)**. Proton bonding to Indium is observed to facilitate elimination of organic functional groups, **Figure 2(d)**.

More in general, it is observed that indium exhibits relatively high affinity for protons. An example is provided in **Figure 8(b)**, in which the Indium atom of a $H_3CIn{=}CH_2$ molecule takes a proton from the vicinal methyl group, thus leading to formation of a $\ddot{C}H_2$ double radical and $HIn{=}CH_2$ molecule, **Figure 8(c)**. It is also worth noting that, due to its larger size (extended valence electron shells, with $4d^{10}5s^{2}5p^{1}$ configuration), indium is a softer ion in comparison to gallium or aluminum. Thus, indium can relatively easily charge as $In^{-}$ both as isolated anion or in a molecule. This explains the stronger propensity of In (in relation to Al) to form indium-hydride functional groups in a molecule, **Figure 2(c, d)** or form individual InH admolecules, **Figure 4(e, f),** and gas molecules, **Figure 2(f)**. Accordingly, InH molecules are prone to or donate protons, **Figure 4(g, h)**, or dissociate into $In^{-}$ ad-anion and $H^{+}$ adproton on surface, **Figure 4(b, c)** and **Figure 8(h, i)**. Analogous arguments suggest that In-C bonds are weaker than Al-C bonds. Plausible implications of different In vs Al (Ref. [11]) reactivity with carbon are (i) relatively quick stripping of the TMIn molecule of its methyl groups, see **Figures 10** and **11**, which contributes to formation of hydrocarbons in the gas phase; (ii) more rare formation of new In-C bonds in comparison to formation of new Al-C bonds observed previously (see figure 6 in Ref. [11]).

Before conclusions, we briefly discuss the results of our simulations in relation to previous experimental observations of TMIn pyrolysis in a hot-wall flow-tube reactor (see Ref. [23]). The experiments were carried out in various chemical environments and temperatures ranging between 573 and 723 K. Consistent with our results, Ref. [23] reports relatively large concentrations of $CH_4$ and $C_2H_6$ hydrocarbons as final products of TMIn decomposition (a summary of reaction products recorded during AIMD simulations is given in **Table 1**). However, the experiments indicated that formation of methane is considerably less frequent than formation of $C_2H_6$ when the reactor does not contain $H_2$ gas. Despite absence of $H_2$, our AIMD simulations of TMIn on free-standing graphene described in the *Supplementary Material* show formation of methane in 3 out of 3, and of ethane and in 2 out of 3, cases (see **Table 1**). The discrepancy with experiments may be due to much higher temperature used in AIMD simulations. Among other relevant similarities with experimental results, we highlight the frequent formation of methyl radical groups as well as of longer chain radicals $C_2H_7$ and $C_3H_7$. In addition, also in experiments, TMIn exhibited higher reactivity in presence of $H_2$, thus confirming chemical affinity between In and hydrogen species.

**Table 1**. Summary of stable and intermediate reaction products of TMIn in a $H_2$ gas on graphene/SiC(0001) at 3500 K and TMIn reactions on free-standing graphene at 4300 K (simulations described in the *Supplementary Material*).

| | Simulation | Stable or intermediate products | | | | | | | | | |
|---|---|---|---|---|---|---|---|---|---|---|---|
| | | In | InH | $InCH_3$ | $CH_4$ | $C_2H_6$ | $C_3H_8$ | HC=CH | $H_2$ | $H^+$ | H |
| TMIn + $H_2$ @ graphene/SiC(0001) 3500 K | #1 | x | x | | | | x | | x | x | |
| | #2 | x | x | x | x | x | | x | | x | |
| | #3 | x | | x | x | | | | | | x |
| | #4 | | | x | | | | | | | |
| TMIn @ graphene | #S1 | x | | x | x | | | x | x | | |
| 4300 K | #S2 | x | x | x | x | x | | x | | x | |
| | #S3 | x | x | x | x | x | | x | x | x | x |

## 5. CONCLUSIONS

We carry out ab initio molecular dynamics simulations of TMIn on free-standing graphene and on graphene supported by SiC in $H_2$ atmosphere. The simulations reveal atomistic pathways for TMIn transformations which ultimately lead to formation of isolated In or InH species on graphene. More specifically, we clarify the role played by $H_2$ on promoting TMIn reactions, show that collisions of TMIn with the graphene surface can activate sequential elimination of methyl radical groups, and illustrate mechanisms for formation of hydrocarbons as methane, ethane, ethene, and propane. More in general, our results are of relevance to interpret or guide experiments that involve MOCVD of TMIn on graphene, or possibly other surfaces.

**Corresponding Author**

*Anelia Kakanakova-Georgieva

Department of Physics, Chemistry and Biology (IFM), Linköping University, 581 83 Linköping, Sweden.

E-mail: anelia.kakanakova@liu.se

## ACKNOWLEDGMENTS

All simulations were carried out using the resources provided by the Swedish National Infrastructure for Computing (SNIC) – partially funded by the Swedish Research Council through Grant Agreement Nº VR-2015-04630 – on the Clusters located at the National Supercomputer Centre (NSC) in Linköping, the Center for High Performance Computing (PDC) in Stockholm, and at the High Performance Computing Center North (HPC2N) in Umeå, Sweden. D.G.S. gratefully acknowledges financial support from the Swedish Research Council (VR) through Grant Nº VR-2021-04426 and the Competence Center Functional Nanoscale Materials (FunMat-II) (Vinnova Grant No. 2016–05156). A.K.-G. and G.K.G. acknowledge support for this work from the Swedish Research Council (VR) through project VR 2017-04071. RF and A.K.-G. acknowledge support from The Swedish Foundation for International Cooperation in Research and Higher Education STINT, project IB2018-7520.

## REFERENCES

[1] www.nobelprize.org/prizes/physics/2014.

[2] Maria Stella Prete, A. M. Conte, P. Gori, F. Bechstedt, O. Pulci, *Tunable electronic properties of two-dimensional nitrides for light harvesting heterostructures*, Appl. Phys. Lett. 110 (2017) 012103.

[3] V. Wang, Z. Q. Wu, Y. Kawazoe, W. T. Geng, *Tunable Band Gaps of* $In_xGa_{1-x}N$ *Alloys: From Bulk to Two-Dimensional Limit*, J. Phys. Chem. C 122 (2018) 6930.

[4] A. Kakanakova-Georgieva, F. Giannazzo, G. Nicotra, I. Cora, G. K. Gueorguiev, P.O.Å. Persson, B. Pécz, *Material proposal for 2D indium oxide*, Appl. Surf. Sci. 548 (2021) 149275.

[5] R. B. dos Santos, R. Rivelino, G. K. Gueorguiev, A. Kakanakova-Georgieva, *Exploring 2D structures of indium oxide of different stoichiometry*, CrystEngComm 23 (2021) 6661.

[6] Y. Guo, L. Ma, K. Mao, M. Ju, Y. Bai, J. Zhao, X. C. Zeng, *Eighteen functional monolayer metal oxides: wide bandgap semiconductors with superior oxidation resistance and ultrahigh carrier mobility*, Nanoscale Horiz. 4 (2019) 592.

[7] B. Pécz, G. Nicotra, F. Giannazzo, R. Yakimova, A. Koos, A. Kakanakova-Georgieva, *Indium Nitride at the 2D Limit*, Adv. Mater. 33 (2021) 2006660.

[8] www.nobelprize.org/prizes/physics/2000.

[9] Z. Y. Al Balushi, K. Wang, R. K. Ghosh, R. A. Vilá, S. M. Eichfeld, J. D. Caldwell, X. Qin, Y.-C. Lin, P. A. DeSario, G. Stone, S. Subramanian, D. F. Paul, R. M. Wallace, S. Datta, J. M. Redwing, J. A. Robinson, *Two-dimensional gallium nitride realized via graphene encapsulation*, Nature Materials 15 (2016) 1166.

[10] A. Kakanakova-Georgieva, G.K. Gueorguiev, D.G. Sangiovanni, N. Suwannaharn, I.G. Ivanov, I. Cora, B. Pécz, G. Nicotra, F. Giannazzo, *Nanoscale phenomena ruling deposition and intercalation of AlN at the graphene/SiC interface*, Nanoscale 12 (2020) 19470.

[11] D. G. Sangiovanni, G. K. Gueorguiev, A. Kakanakova-Georgieva, *Ab initio molecular dynamics of atomic-scale surface reactions: insights into metal organic chemical vapor deposition of AlN on graphene*, Phys. Chem. Chem. Phys. 20 (2018) 17751.

[12] N. Nayir, M. Y. Sengul, A. L. Costine, P. Reinke, S. Rajabpour, A. Bansal, A. Kozhakhmetov, J. Robinson, J. M. Redwing, A. van Duin, *Atomic-scale probing of defect-assisted Ga intercalation through graphene using ReaxFF Molecular Dynamics Simulations*, Carbon 190 82022) 276.

[13] G. Kresse, J. Hafner, J. *Ab initio molecular dynamics for liquid metals*, Physical Review B 47 (1993) 558.

[14] D. M. Ceperley, B. J. Alder, *Ground-state of the electron gas by a stochaistic method*, Phys. Rev. Lett. 45 (1980) 566; J. P. Perdew, A. Zunger, *Self-interaction correction to density-functional approximations for many-electron systems*, Phys. Rev. B 23 (1981) 5048.

[15] P. E. Blöchl, *Projector augmented-wave method*, Phys. Rev. B 50 (1994) 17953.

[16] S. Grimme, *Semiempirical GGA-type density functional constructed with a long-range dispersion correction*, Journal of Computational Chemistry 27 (2006) 1787.

[17] D. G. Sangiovanni, O. Hellman, B. Alling, I. A. Abrikosov, *Efficient and accurate determination of lattice-vacancy diffusion coefficients via non equilibrium ab initio molecular dynamics*, Phys. Rev. B 93 (2016) 094305.

[18] D. G. Sangiovanni, A. B. Mei, L. Hultman, V. Chirita, I. Petrov, J. E. Greene, *Ab Initio Molecular Dynamics Simulations of Nitrogen/VN(001) Surface Reactions: Vacancy-Catalyzed N-2 Dissociative Chemisorption, N Adatom Migration, and N-2 Desorption*, J. Phys. Chem. C 120 (2016) 12503.

[19] W. Humphrey, A. Dalke, K. Schulten, *VMD: Visual molecular dynamics*, Journal of Molecular Graphics & Modelling 14 (1996) 33.

[20] A. J. Cohen, P. Mori-Sanchez, W. T. Yang, *Insights into current limitations of density functional theory*, Science 321 (2008) 792, Editorial Material.

[21] S. Chen, P. A. Thiel, E. Conrad, M. C. Tringides, *Growth and stability of Pb intercalated phases under graphene on SiC*, Phys. Rev. Mater. 4 (2020) 124005.

[22] Y. Han, J. W. Evans, M. C. Tringides, *Dy adsorption on and intercalation under graphene on 6H-SiC (0001) surface from first-principles calculations*, Phys. Rev. Mater. 5 (2021) 074004.

[23] A. H. McDaniel, M. D. Allendorf, *Autocatalytic Behavior of Trimethylindium during Thermal Decomposition*, Chem. Mater. 12 (2000) 450.